\documentclass[a4paper,aps,prl,showpacs,twocolumn,nofootinbib]{revtex4-1}

\pdfoutput=1

\usepackage{amsmath,amssymb,graphicx}
\usepackage{color}
\usepackage[normalem]{ulem}

\newcommand{\be}{\begin{equation}}
\newcommand{\ee}{\end{equation}}
\newcommand{\bea}{\begin{eqnarray}}
\newcommand{\eea}{\end{eqnarray}}
\newcommand{\bean}{\begin{eqnarray*}}
\newcommand{\eean}{\end{eqnarray*}}

\newcommand{\BB}{{\cal B}}

\newcommand{\MM}{{\cal M}}

\newcommand{\PP}{{\cal P}}

\newcommand{\de}{\delta}

\newcommand{\gsim}{\stackrel{>}{\sim}}

\def\id{{\rm 1\kern -2.5pt I}}

\definecolor{dgreen}{rgb}{0,0.6,0} 

\begin{document}

\title{Impact of next-to-leading order contributions to CMB lensing}

\author{Giovanni Marozzi$^1$,  Giuseppe Fanizza$^2$,  Enea Di Dio$^{3,4,5}$ and Ruth Durrer$^6$,}

\affiliation{$^1$Centro Brasileiro de Pesquisas F\'{\i}sicas, Rua
  Dr. Xavier Sigaud 150, Urca,  CEP 22290-180, Rio de Janeiro, Brazil\\
  $^2$Center for Theoretical Astrophysics and Cosmology, Institute for Computational Science, University of Z\"urich, CH-8057 Z\"urich, Switzerland\\
$^3$INAF - Osservatorio Astronomico di Trieste, Via
  G. B. Tiepolo 11, I-34143 Trieste, Italy \\
  $^4$SISSA- International School for Advanced Studies, Via Bonomea 265, 34136 Trieste, Italy \\
  $^5$INFN - National Institute for Nuclear Physics,
via Valerio 2, I-34127 Trieste, Italy \\
$^{6}$Universit\'e de Gen\`eve, D\'epartement de Physique Th\'eorique and CAP,
24 quai Ernest-Ansermet, CH-1211 Gen\`eve 4, Switzerland}

\date{\today}

\begin{abstract} 

In this Letter we study the impact on cosmological parameter estimation, from present and future surveys, 
due to  lensing corrections on CMB temperature and polarization anisotropies beyond leading order. 
In particular, we show how post-Born corrections, LSS effects and the correction due to the change in the polarization direction between the emission at the source and the detection at the observer, are non-negligible in the determination
of the polarization spectra. They have to be taken into account for an accurate estimation of  cosmological 
parameters sensitive to or even based on these spectra.  We study in detail the impact of higher order lensing on the determination of the tensor-to-scalar ratio $r$ and on the estimation of the effective number of relativistic species $N_\text{eff}$.
We find that neglecting higher order lensing terms can lead to misinterpreting these corrections as a primordial tensor-to-scalar ratio of about $\mathcal{O}(10^{-3})$. 
Furthermore, it leads to a shift of the parameter 
$N_\text{eff}$ by nearly two sigma considering  the level of accuracy aimed by future S4 surveys.

\end{abstract}

\pacs{98.80.-k, 98.80.Es}

\maketitle

{\it Introduction.} In this Letter we discuss the effects of lensing of the temperature and polarization
anisotropies of the Cosmic Microwave Background radiation (CMB) beyond leading order. It is well known that gravitational lensing affects not only the CMB temperature fluctuations but also the polarization~\cite{Seljak:1995ve,Hirata:2003ka,Lewis:2006fu,RuthBook}. 
It actually transforms E-polarization (gradient type polarization) into B-polarization (curl-type polarization) which has to be subtracted from the primordial $B$-mode before it can be used to constrain the tensor-to-scalar ratio. This lensing-induced B-mode has already been measured~\cite{Ade:2013gez,Ade:2015tva,Ade:2015nch}.  The CMB, the most precise cosmological dataset, therefore not only allows us to determine with high precision the fluctuations of the space-time geometry and the matter distribution at the time of last scattering, but via lensing, it allows also to determine the (integrated) fluctuations in the geometry at lower redshifts, mainly due to the growth of matter density perturbations.  One of the important goals of future CMB experiments 
is to measure lensing out to the CMB with high precision and generate precise lensing maps~\cite{Ade:2013gez,
Story:2014hni,
Ade:2015zua,
Array:2016afx,
Sherwin:2016tyf,
Abazajian:2016yjj,DiValentino:2016foa}. 
 
 The effects of lensing in the CMB are quite substantial. For $\ell\sim 800$
 its effect on temperature anisotropies is about 1\% and it raises to 10\% by $\ell\sim 2200$. For $\ell>5000$ it dominates over the primordial signal.  Present analysis is mainly based on the first order perturbations of the CMB from lensing. The Taylor series in the first order deflection angle, however, has to be  resummed. This can be performed exactly assuming a Gaussian deflection angle~\cite{Lewis:2006fu,RuthBook}. It turns out that this resummation is quite important, e.g. for cosmological parameter estimation with CAMB~\cite{Lewis:1999bs} or CLASS~\cite{Lesgourgues:2011re,Blas:2011rf}, already for present experiments like Planck~\cite{Planck:2015xua}.  It is therefore reasonable to ask for a systematic next-order calculation including especially also non-Gaussian contributions (e.g. from an odd number of deflection angles) which are present at higher order.
This has been attempted in~\cite{Marozzi:2016uob} for the temperature and in \cite{Hagstotz:2014qea,Pratten:2016dsm,Lewis:2016tuj} for both the temperature and the polarization.  
In particular, in \cite{Marozzi:2016uob} the contribution from the non-Gaussian 
post-Born terms was evaluated for the first time and its dominance over other post-Born corrections has been shown.
Nevertheless, all the new contributions beyond the leading order lensing 
of the CMB temperature anisotropies, 
as found in these references, are quite small.

In this Letter, starting from the results of \cite{CMBlensing2}, we study the effect of higher order lensing of CMB temperature and polarization on the evaluation of cosmological parameters.
In \cite{CMBlensing2} the beyond leading order lensing of the CMB polarization
was evaluated in full generality, going beyond the results given in \cite{Hagstotz:2014qea,Pratten:2016dsm,Lewis:2016tuj}.
Especially,  a new effect has been considered in \cite{CMBlensing2}, which was overlooked previously, and which actually dominates the next-to-leading order correction for B-polarization.
The physical origin of this new effect is quite simple: the CMB polarization tensor is parallel transported along a perturbed lightlike geodesic. Therefore, we see the photons in a direction which differs from the one into which they have been emitted. But parallel transport of a tensor also rotates its principal axes. It is well known that for purely scalar lensing this rotation vanishes, but if the lens map has a curl component~\cite{Dai:2013nda}, this leads to a rotation of the polarization direction.
It also induces a small additional deflection, but it has been shown in~\cite{Lewis:2016tuj} that this is very subdominant. Here we show how the effect coming from the rotation of the polarization axis, together with the other 
higher order contributions, can have a substantial impact on the determination of cosmological parameters.

The new effect described above is
especially important since the curl-lensing transforms E-polarization into B-polarization and,
together with the other higher order lensing effects, 
modifies the  B-modes from lensing by up to about 0.1\% for $\ell<1500$
and enhances them by nearly 2\% for $\ell\simeq 3500$  (see \cite{CMBlensing2}). 
These contributions are relevant when performing the lensing subtraction of B-modes in order to recover the primordial signal  from inflationary gravitational waves. If discovered, this signal, and especially the tensor-to-scalar ratio $r$, will provide the energy scale of inflation and allow us a glimpse to the physics at the highest energies ever observed, many orders of magnitude higher than the energy achieved at the LHC at CERN.
It is therefore very important to identify all possible foregrounds which might hamper the detection of~$r$.
 
 But even independent of primordial B-modes, the detection of the lensing curl mode in B-polarization will be a formidable test of General Relativity (GR)
 as it is a measurement of gravitational `frame dragging', an effect detected for the first time with Gravity Probe B~\cite{Everitt:2011hp} from the gravitational field of the rotating earth. Seeing the lensing-curl rotation of B-mode polarization tests frame dragging on cosmological scales. This shows once more, that cosmology offers many opportunities to actually test GR on large scales.
 
 Higher order lensing corrections of the polarization spectra are not just important for the 
 determination of the tensor-to-scalar ratio. As we show in this Letter, they
  are also relevant for cosmological parameter estimation, especially at the level of 
 precision aimed in future survey~\cite{Abazajian:2016yjj}.

{\it Methods.} Here we present the basic ideas and sketch the derivation of the results obtained 
in~\cite{CMBlensing2}. 
A much more detailed description is provided in~\cite{CMBlensing2}, where the analytical expressions are presented and a comparison of the magnitude of the effects with cosmic variance 
is performed.

When observing a scalar quantity in the sky (e.g. the temperature anisotropy) in direction 
$\boldsymbol{\theta}$,
we have to take into account that   this direction is deflected by lensing and the true direction of 
the source
is $\boldsymbol{\theta} +\delta\boldsymbol{\theta}$. The observed temperature anisotropy, let us call it $\tilde\MM$, is therefore related to the anisotropy on the last scattering surface, which we denote $\MM$, by
 \be
 \tilde\MM(\boldsymbol{\theta}) = \MM(\boldsymbol{\theta} +\delta\boldsymbol{\theta}) \,.
 \ee
 The deflection is caused by foreground structures and can be considered as uncorrelated with the temperature anisotropy. Since $\langle\de\boldsymbol{\theta}\rangle=0$, a systematic expansion yielding the leading and next-to-leading corrections to the power spectrum has to go up to third order in the deflection angle and must include terms with up to four angles,
 \vspace{-2mm}
\bea
\!\!\!\!\!\!\!\! \tilde\MM(\boldsymbol{\theta}) &=& \MM(\boldsymbol{\theta}) +\sum_{i=1}^3\delta\theta^{a\,(i)}\nabla_a\MM  
 \nonumber \\    &&
+\frac{1}{2}\sum_{i+j\leq 4}\delta\theta^{a\,(i)}\delta\theta^{b\,(i)}\nabla_a\nabla_b\MM  \nonumber \\
 &&  +\frac{1}{6}\sum_{i+j+k\leq 4}\delta\theta^{a\,(i)}\delta\theta^{b\,(j)}\delta\theta^{c\,(k)}\nabla_a\nabla_b\nabla_c\MM 
\nonumber \\
 &&
 +\frac{1}{24}\delta\theta^{a\,(1)}\delta\theta^{b\,(1)}\delta\theta^{c\,(1)}\delta\theta^{d\,(1)}
 \nabla_a\nabla_b\nabla_c\nabla_d\MM \,.
 \label{e:Mlens}
 \eea 
The superscripts in parentheses indicate the order at which the deflection angle has to be taken,  the Latin indices $a$ to $d$ go over the two directions of the sphere and $\nabla_a$ are covariant derivatives on the sphere.
 Introducing the Weyl potential as the mean of the Bardeen potentials~\cite{RuthBook}
 \vspace{-4mm}
 \be
\Phi_W =\frac{1}{2}(\Phi+\Psi)\,,
\nonumber
\ee
the deflection angle can be determined recursively as weighted integral of $\Phi_W$, it is given in \cite{Fanizza:2015swa}.

To obtain the spectra one then converts Eq.~\eqref{e:Mlens} into spherical harmonic space. Details are presented in~\cite{Marozzi:2016uob}.

At second order the Jacobi map, which describes the evolution of the geodesic deviation vector, not only contains expansion and shear but also a rotation due to the induced vector mode.  This represents the rotation $\beta$ of the geodesically transported Sachs basis of the screen. The CMB
 polarization tensor given in terms of the Stokes parameters $Q$ and $U$ as  $\PP_\pm=Q\pm iU$, which has helicity $\pm2$ transforms with $e^{\pm2i\beta}$ under this rotation. It is well known that a rotation of the polarization pattern transforms E-modes into B-modes and vice versa~\cite{RuthBook,Kamionkowski:2008fp}. 
To take this effect into account up to next-to-leading order, we need in principle  the rotation angle up to 4th order. 
But  $\beta^{(0)}=0$ and for scalar perturbations also $\beta^{(1)}=0$.
The only terms from rotation in the  
 power spectrum are therefore of the form $\left(\beta^{(2)}\right)^2$ and we can neglect 
 $\beta^{(3)}$ and $\beta^{(4)}$.
 The expectation value of terms containing only one rotation angle have to vanish for parity reasons.
In~\cite{CMBlensing2} we obtain
\bea
   \beta^{(2)}(\boldsymbol{\theta}) &=& 
   2\,\epsilon^{ab}\int_{0}^{r_s}dr\,\frac{r_s-r}{r_s\,r} \nabla_a\nabla^c\Phi_W(r)
   \nonumber \\ 
   & & \times   
   \int_0^rdr_1\,\frac{r-r_1}{r\,r_1}\nabla_b\nabla_c\Phi_W(r_1)
   \,,
\eea
where $\Phi_W(r) \equiv\Phi_W(\eta_0-r, r\boldsymbol{\theta})$.
This rotation angle is identical to the rotation angle in the lens map usually called $\omega$, even if these angles have slightly different physical meanings, as shown in \cite{Dai:2013nda} for first order vector and tensor perturbations. Indeed, 
$\omega$ refers to a bundle of light rays and describes the rotation of an image along the line-of-sight, whereas $\beta$ acts also on a single photon yielding the rotation of a geodesically transported vector. Nevertheless, in \cite{CMBlensing2} a direct evaluation of $\beta$ by computing the Sachs basis, in geodesic light-cone gauge (see \cite{Gasperini:2011us,Fanizza:2013doa,Fanizza:2014baa}), and the straightforward derivation of $\omega$ as the antisymmetric part of the amplification matrix, shows the equality of the two angles.

At next-to-leading order, we have three contributions: 
pure post-Born terms due to higher order deflection angles, the LSS contribution to 
higher order deflection angles (obtained by expanding the Weyl potential beyond linear order), 
and, for the polarization spectra, corrections from the rotation angle $\beta$~\footnote{The contribution from the rotation can be further divided into two parts, one of them is a ``constant'' contribution which just leads to an
overall shift of 
$\Delta C_{\ell}^{XX}/C_\ell^{XX}$ (see \cite{CMBlensing2} for details). 
This part is negligible in cosmological parameters estimation and we do not consider it in this Letter.}.
The general expression for a lensed spectrum can be written as
\be
\tilde C_\ell^{XY} =\tilde C_\ell^{XY\, (1)} + \Delta C_{\ell,pB}^{XY}  + 
\Delta C_{\ell,LSS}^{XY} +  \Delta C_{\ell,\beta}^{XY}  \,,
\label{eq:cltot}
\ee
where $\tilde C_\ell^{XY\, (1)}$ denotes the unlensed $C_\ell^{XY}$ plus the well known resummed correction from the first order
deflection angle \cite{Seljak:1995ve,RuthBook,Lewis:2006fu}.
The post-Born and LSS contributions can be divided into two groups. Following the formalism 
of \cite{Marozzi:2016uob,CMBlensing2}, we have a ``second group'' which involves two point correlation functions of deflection angles 
(first$\times$third or second$\times$second order),
and a ``third group'' which contains three point correlation functions of deflection angles (first$\times$first$\times$second order).
Hereafter, we use Halofit model (see~\cite{Smith:2002dz,Takahashi:2012em}) for the matter power spectrum.
The second group of LSS corrections is then included in the leading terms. Therefore this  contribution is not added separately.
Beyond Halofit, our results can be generalized also to fitting formulae for the bispectrum \cite{GilMarin:2011ik}. These fitting 
formulae can then be used to evaluate the full higher-order effects on the 
bispectrum~\cite{Namikawa:2016jff,Bohm:2016gzt}.

\begin{figure}[ht!]
\includegraphics[width=\linewidth]{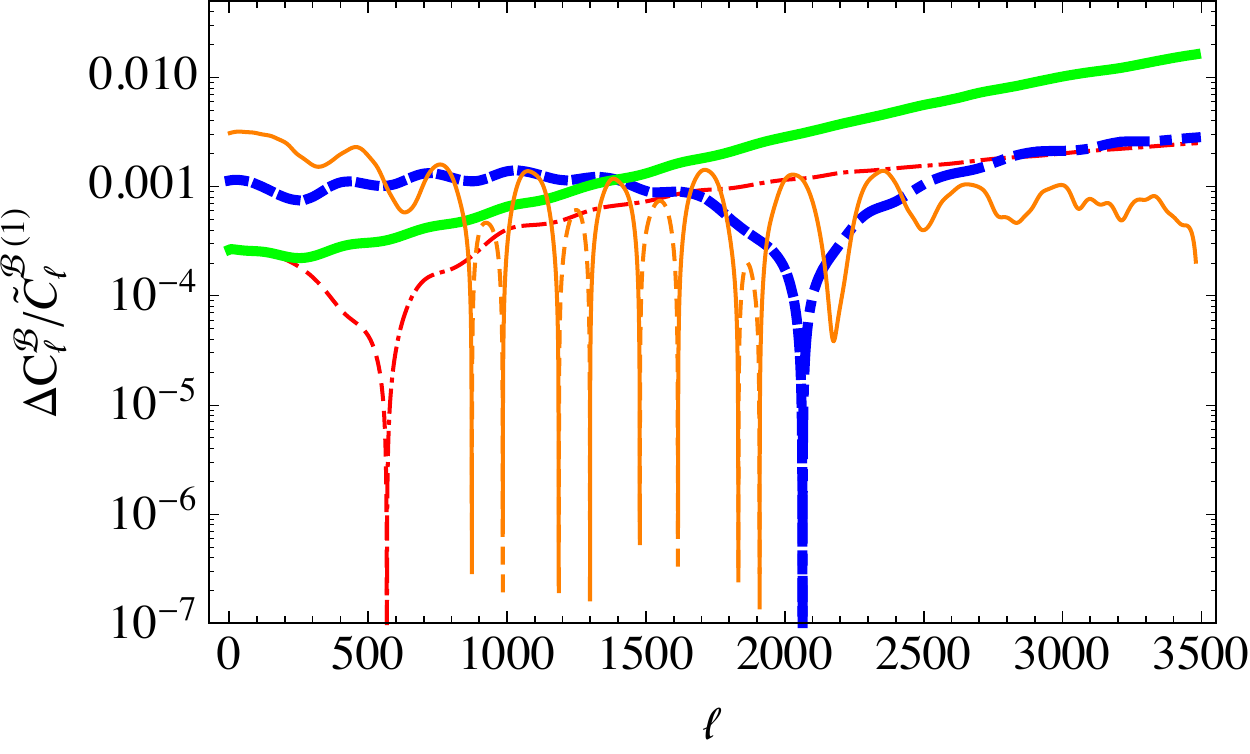}
\caption{Next-to-leading order corrections to B-modes. 
We show the correction from post-Born second group (thin dot-dashed red), post-Born third group (thick dot-dashed blue), LSS (thin orange), 
and the contribution coming from the rotation angle $\beta^{(2)}$ (thick green), negative values are dashed. 
All functions are normalized to the resummed first order deflection angle $\tilde C^{\BB(1)}_\ell$ which is included in standard CMB codes.}
\label{fig:CBB}
\end{figure}

\begin{figure}[ht!]
\centering
\includegraphics[width=\linewidth]{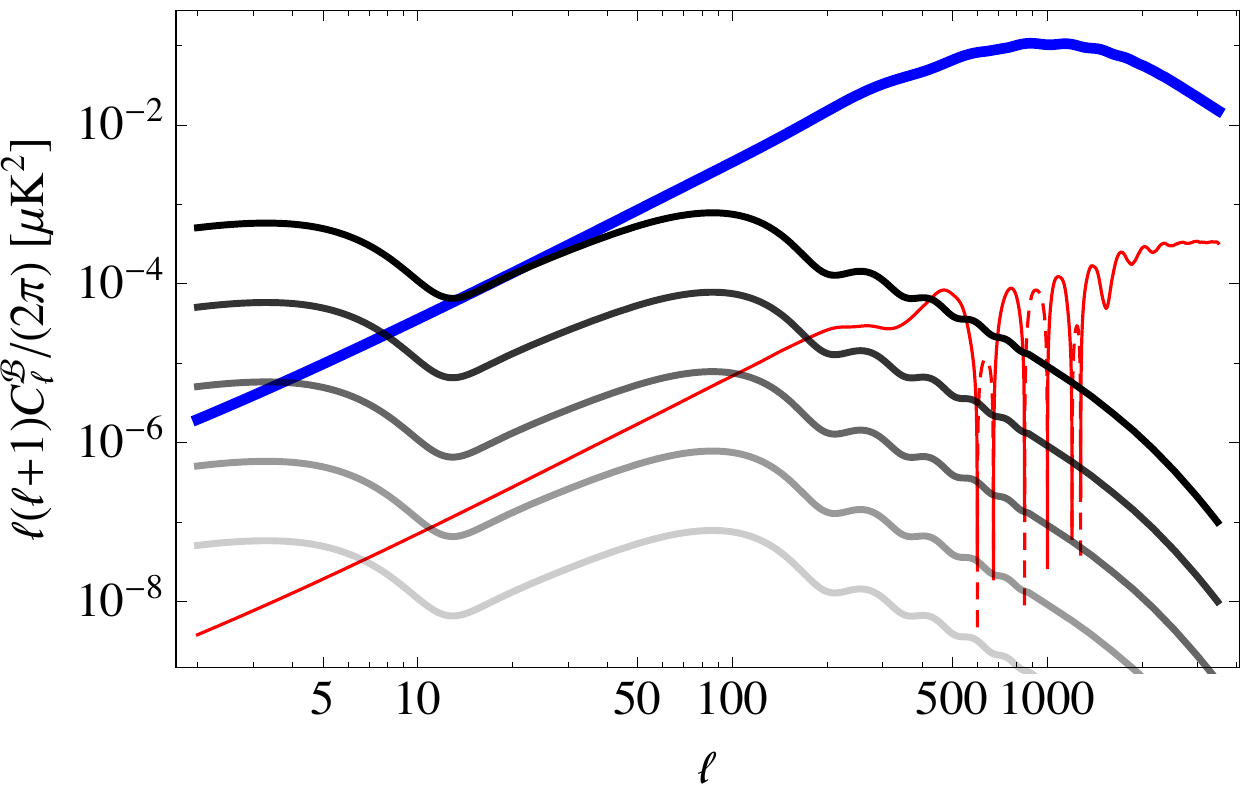}
\caption{The first order resummed lensing correction to the B-modes of the CMB (thick blue line) and the additional correction from the full higher order contribution (post-Born+LSS+rotation, thin red line) are compared. Grey lines refer to primordial B-modes with no lensing for different values of $r$,  $r=\left(10^{-2}, 10^{-3}, 10^{-4}, 10^{-5}, 10^{-6}\right)$ from top to bottom. Negative values are dashed.}
\label{fig:r}
\end{figure}

{\it Results.} In~\cite{CMBlensing2} we have evaluated the corrections summarized in Eq.~\eqref{eq:cltot} to  the temperature and the polarization spectra.
In particular, the new contribution coming from the rotation turns out to be non-negligible for the 
E-mode
spectrum, and 
leading at high $\ell$ for the 
B-mode
spectrum, see \cite{CMBlensing2} for details.
In Fig.~\ref{fig:CBB} we show the most interesting case, i.e.~the ratio of the new terms to the standard (first order resumed) lensing-induced B-mode. 
As the figure clearly shows, the contribution from rotation not only cannot be neglected but actually dominates at $\ell>1500$.
Different higher order contributions dominate in different $\ell$-ranges.
At low $\ell$ the non-Gaussian post-Born contribution and the LSS contribution dominate with different signs and they partially cancel.
This is particularly relevant for $\ell<300$, the range used  to infer the value of $r$, see~e.g.~\cite{Ade:2015tva}.
The importance of including higher order lensing when performing the lensing subtraction of B-modes, in order to recover the primordial signal  from inflationary gravitational waves, is evident from Fig.~\ref{fig:r}.
There we compare the B-mode from lensing with a possible primordial signal for different values of $r$
and including the leading correction and the total higher order contribution (sum of post-Born, LSS and rotation). 
Clearly, higher order lensing correction can be misinterpreted for a primordial signal at the level of $r \simeq 10^{-3}$ or more from $\ell>200$, even if the leading terms are correctly subtracted. Namely, if we neglect this correction we could erroneously 
measure a primordial signal using future surveys which plan to have a level of accuracy which  would be sensitive to 
$r \simeq  {\cal O} (10^{-3})$~\cite{Abazajian:2016yjj}. 
The impact of higher order lensing on the $r$ measurement is lower for $\ell$ up to 
$\ell_{\max}=200$, but it is still at the level of $r \simeq 4.5 \times10^{-4}$ for $\ell=150$ and to measure $r \simeq  10^{-3}$ at $3\sigma$ the higher order lensing correction has to be subtracted.

From Fig.~\ref{fig:CBB} we infer that, at small $\ell$  the correction 
to the lensing contamination can be at the level of $0.1$\%.  While, at high $\ell\gsim 2500$ 
it adds a contribution at the percent level to the lensing power spectrum. It is important to remember that here we show the power spectra, in the amplitude of the fluctuations this contribution is more than 10\%.

\begin{figure}[ht!]
\centering
\includegraphics[width=0.9\linewidth]{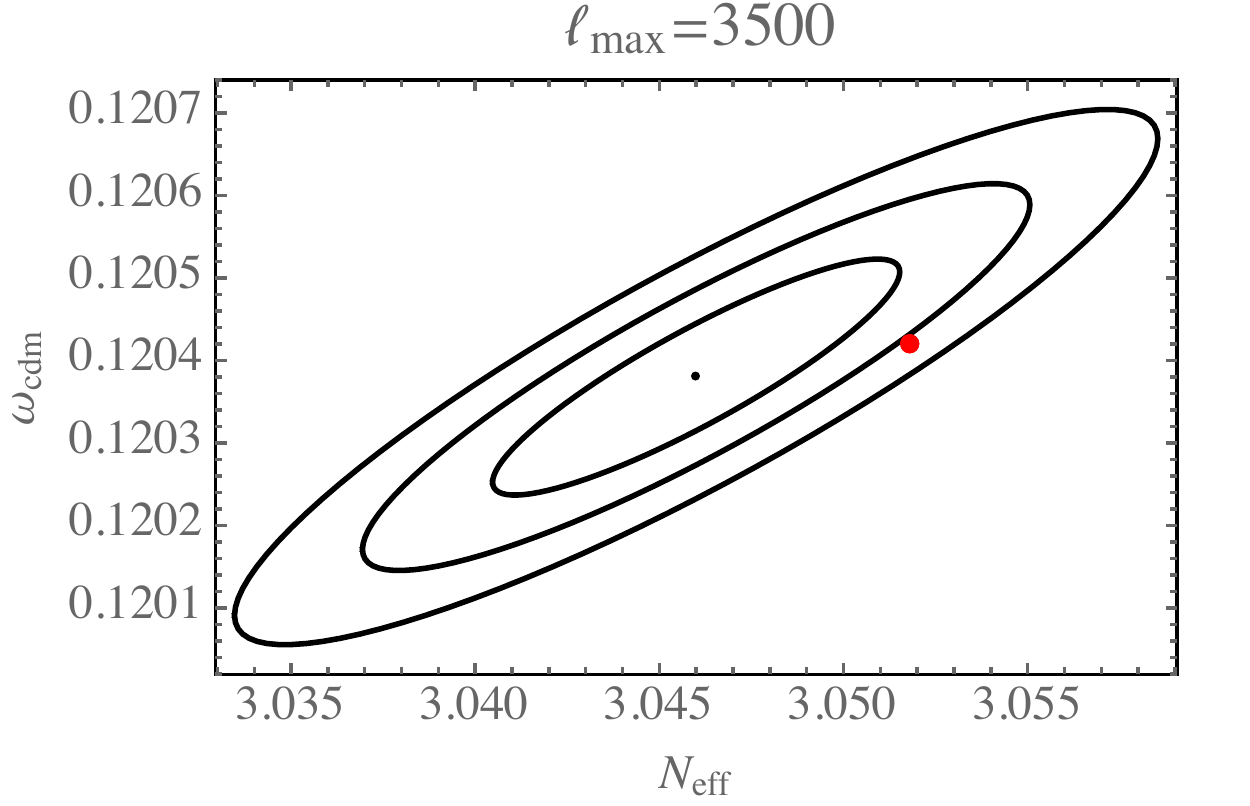}\\
\caption{The bias introduced in cosmological parameter estimation if sub-leading CMB lensing effects are neglected. We consider multipoles up to $\ell_\text{max}=3500$ for TT, EE, TE and $\ell_\text{max}=1500$ for BB spectra, for an ideal  cosmic variance limited survey with sky coverage $f_{\rm sky}=0.75$ for TT, EE, TE  and $f_{\rm sky}=0.5$ for BB spectra, keeping
 all the parameters not shown fixed at their fiducial values.
}
\label{fig:Fisher}
\end{figure}

Let us also consider another example to show the importance of the higher order lensing correction in the determination of cosmological parameters with  future CMB survey.
We discuss the effect of higher order lensing for the determination 
of the effective number of relativistic species $N_\text{eff}$.
In Fig.~\ref{fig:Fisher}  we show the
theoretical bias introduced in cosmological parameter estimation if higher order CMB lensing contributions are neglected. 
Considering as parameters the effective number of relativistic species and the dark matter density parameter, $\omega_\text{cdm}=h^2\Omega_\text{cdm}$,  and keeping all the other parameters  fixed at their fiducial values, we have performed a Fisher matrix analysis considering temperature, 
E-modes
and TE  correlation power spectra up to $\ell=3500$ for a cosmic variance limited survey with sky coverage $f_\text{sky}=0.75$ for TT, EE, TE and $f_\text{sky}=0.5$ for BB spectra.
Neglecting higher order 
lensing contributions leads to a non-negligible shift (about 2 sigma) of the measured parameters with respect to their true value. The shift becomes even larger for the ideal case of full sky coverage, i.e. well beyond 2 sigma.
In performing this analysis we have in mind a S4 CMB experiment with a conservative value of the
attainable upper bound of $C_\ell\approx 2.5\times10^{-7}\,\mu K^2$ \cite{Abazajian:2016yjj}. 
In this case we can reach $\ell \simeq 3500$ in the E-mode spectrum (and even higher values for the temperature and TE power spectra). 
 We also consider B-modes up to $\ell \simeq 1500$, however they carry less information. 
Furthermore, considering that using a Fisher matrix technique approximates the spectra as Gaussian fields, which underestimates the contribution to the covariance induced by the B-mode (see e.g.~\cite{Smith:2006nk}), the true
constraining power of B-mode is probably even smaller.
Keeping the other cosmological parameters fixed at their fiducial values means that we  assume that they are determined with good accuracy by other cosmological probes. When marginalising also over $H_0$ the shift reduces to somewhat more than $1$-sigma, while marginalising over all
 cosmological parameters reduces the shift  below $1$-sigma.

These findings demonstrate the importance of higher order lensing for a very accurate determination of parameters like 
the effective number of relativistic species. 
Moreover, if we consider the optimistic upper bound of $C_\ell\approx 10^{-8}\,\mu K^2$ achievable by 
S4 CMB experiments, we can 
go to larger values of $\ell$ 
and the shift becomes even more significant.
In this optimistic case the power spectra are measured roughly at the cosmic variance accuracy up to the values of $\ell_\text{max}$ considered in the Fisher matrix analysis.

Clearly, the Fisher matrix is an approximation and a full 
Monte Carlo analysis would be more reliable. The Fisher matrix 
can only be trusted when the parameter shift is small (when it is large it can differ significantly from a Monte Carlo result \cite{Cardona:2016qxn}). However, the present analysis is a proof of principle demonstrating that cosmological parameters inferred from S4-precision CMB data can shift by more than 1 sigma due to higher order lensing terms.

{\it Conclusions.} In this Letter we have shown that next-to-leading order lensing corrections to CMB temperature and polarization 
spectra can be crucial in cosmological parameter estimation from CMB data, and 
cannot be neglected in future CMB experiments.

In particular, lensing at higher order significantly modifies B-mode polarization. Even though the signal is quite small, it has to be included for lensing reconstruction of primordial B-modes.
If not, one erroneously measures a primordial tensor-to-scalar ratio at the level of $10^{-3}$ in the multipole range $200<\ell<300$ (although at larger scales,  lower $\ell$, the impact is smaller). 

On smaller scales, $\ell>1500$ the new lensing correction is rather large and it is dominated by the rotation induced by the curl part of the lensing potential. This signal is purely  second order in perturbation theory and therefore non-Gaussian, which can help to identify it. Measuring this spin-1 signal will be possible with S4-precision experiments (see \cite{Fabbian:2017wfp}) and it represents a measure of
 `frame dragging' in the Universe. Finding the best way to measure $\beta$ in an interesting open problem. The proposal of Ref.~\cite{Kamionkowski:2008fp} does not work in this case since the random rotation angle $\beta$ is virtually  uncorrelated with the CMB anisotropies and polarization and does not lead to non-vanishing $EB$ or $TB$ cross spectra.

To conclude, not only the correction of B-modes are important but, as we have shown in Fig.  \ref{fig:Fisher}, the 
higher order lensing corrections to $TE$ and $EE$
have to be  included to measure, 
for example, the effective number of relativistic species at the level of accuracy aimed at by future surveys, see \cite{Abazajian:2016yjj}. 
The higher order contribution to the E-spectrum is partly degenerate with the modifications from a change in the effective number of relativistic species.
Neglecting the next-to-leading order corrections to E polarization lead to a shift of the cosmological parameter 
 $N_\text{eff}$ of nearly two sigma when considering multipoles up to $\ell=3500$, see Fig.~\ref{fig:Fisher}.

We  thank Giulio Fabbian and Tom Crawford for helpful discussions.
GM wishes to thank CNPq for financial support.  GF is supported by a Consolidator
Grant of the European Research Council (ERC-2015-CoG
grant 680886). 
ED is supported by the ERC Starting Grant cosmoIGM and by INFN/PD51 INDARK grant.
RD acknowledges support from the Swiss National Science Foundation.

\bibliographystyle{apsrev4-1}
\bibliography{biblio_CMBlensing}

\end{document}